\documentstyle[preprint,aps]{revtex}
\draft
\begin{document}
\title{Peltier effect in NS microcontacts }
\author{A. Bardas and D. Averin }
\address{Department of Physics, SUNY Stony Brook, NY 11794}
\maketitle

\begin{abstract}
We have calculated the heat current in the normal
metal/insulator/ superconductor contacts with arbitrary
transparency of the insulator barrier. In the tunneling limit
(small transparencies), the heat flow out of the normal metal
reaches its maximum at temperature $T\simeq 0.3\Delta$. At higher
values of transparency, the interplay between single-particle
tunneling and Andreev reflection determines optimum transparency
which maximizes the density of heat flow out of the normal metal.
In clean contacts, the optimum transparency is about $0.1$ at
$T=0.3\Delta$ and decreases with temperature roughly as
$(T/\Delta)^{3/2}$. In disordered contacts, disorder
enhances Andreev reflection and shifts the optimum point
towards smaller transparencies. The optimal ratio of the barrier
resistance to the resistance of the normal electrode is
$R_N/R_T \simeq 0.01$ at $T=0.3 \Delta$ and decreases with
temperature similarly to clean contacts. For disordered
contacts we also plot current-voltage characteristics for
arbitrary values of the ratio $R_N/R_T$.
\end{abstract}
%\pacs{PACS numbers: 73.50.Lw, 74.50.+r, 74.80.Fp }
\narrowtext
\newpage

\section{Introduction}
\label{sec:intr}

\vspace{1ex}

The flow of electric current in the normal
metal/insulator/superconductor (NIS) contacts is accompanied by
the heat transfer from the normal metal into the superconductor.
This principle can be applied to the refrigeration of electrons in
the normal metal \cite{b1}. (Implicitly, the same principle is
used in the enhancement of the superconductivity in the SIS'IS
structures -- see \cite{b2,b3} and references therein.) The
mechanism of the heat transfer in the NIS contacts is the same as
that of the well-known Peltier effect in metal/semiconductor
contacts -- see, e.g., \cite{b4}.
Due to the energy gap in the superconductor, electrons with
higher energies (above the gap) are removed from the normal metal
more effectively than those with lower energies. This makes the
electron energy distribution sharper, thus decreasing the
effective temperature of electron gas in the normal metal.

In the limit of very small transparencies $D$ of the insulator
barrier, when the mechanism of electron transport across the
barrier is the single-particle tunneling, the magnitude of
heat flow out of the normal metal increases with transparency.
However, at larger transparencies, coherent two-electron tunneling
(``Andreev reflection'') starts to dominate electron transport and
suppress the heat flow. This occurs because in the Andreev reflection
electrons with all energies, including those inside the energy gap,
are removed from the normal metal. Below we study the crossover
between the two regimes and find the optimum transparency which
maximizes the heat flow through a unit area of the NS interface. The
calculations are carried out in the two cases of contacts with
clean and disordered electrodes. It is shown that, in accordance
with the general understanding that disorder enhances Andreev
reflection\cite{b5,b6,b7}, in disordered contacts the optimum
transparency is shifted towards smaller transparencies with
increasing disorder.

\section{Clean contacts}  \label{sec:clean}

\vspace{1ex}

The model of NIS contact we consider is a constriction between
normal metal and superconductor with characteristic dimensions $d$
that are much smaller than the coherence length $\xi$ and inelastic
scattering length in the electrodes. Because of the condition
$d\ll \xi$, we can neglect variation of the superconducting order
parameter $\Delta$ in the vicinity of the constriction and solve
the problem, assuming that $\Delta$ is constant in space up to
the NS boundary and is equal to its equilibrium value inside
the superconductor.

The properties of such a constriction depend strongly on the
relation between $d$ and elastic scattering length $\ell$ near
the junction. In the clean limit ($d\ll \ell$), electron motion
in the constriction is naturally decomposed into several
uncoupled transverse modes (provided that the NS interface
is smooth and conserves transverse momentum). In this
case one can follow Blonder, Klapwijk, and Tinkham \cite{b8} and
solve the Bogolyubov-de Gennes (BdG) equations independently for
each transverse mode. The
basic result of such a solution (for details see, e.g., the review
\cite{b9}) is contained in the probabilities of normal ($B$) and
Andreev ($A$) reflection from the NS interface as functions of
the quasiparticle energy $\epsilon$:
\begin{equation}
A(\epsilon )= \frac{D^2 |a(\epsilon)|^4 }{1 -2R \mbox{Re}
a^2(\epsilon) +R^2|a(\epsilon)|^4 }\, , \;\;\;
B(\epsilon )= \frac{R(1-2\mbox{Re}a^2(\epsilon)+ |a(\epsilon)|^4)
}{1-2R \mbox{Re}a^2(\epsilon) +R^2|a(\epsilon)|^4} \, .
\label{1} \end{equation}
Here $D,R$ are transmission and reflection probabilities of the
insulator barrier, $D+R=1$, which are assumed to be independent of
energy on the energy scale given by $\Delta$, and $a(\epsilon)$ is
the amplitude of Andreev reflection from the ideal NS interface
with $D=1$:
\begin{equation}
a(\epsilon)=\frac{1}{\Delta} \left\{ \begin{array}{ll}
\epsilon- \mbox{sign}(\epsilon) (\epsilon^2- \Delta^2)^{1/2}\, ,
\;\;\; & \mid \epsilon \mid > \Delta\, , \\
\epsilon-i(\Delta^2 -\epsilon^2)^{1/2}\, , \;\;\; & \mid \epsilon
\mid < \Delta\, . \end{array} \right.
\label{2} \end{equation}

Using the reflection probabilities (\ref{1}), we can write
the balance equation for energy distribution of electrons moving
to and from the NS interface \cite{b8}. Energy distribution of
electrons $f^+(\epsilon)$ that move from the bulk of the normal
metal to the NS interface is the equilibrium Fermi distribution
shifted by $eV$, $f^+(\epsilon)=f(\epsilon-eV)$. Electrons moving
from the interface into the normal metal are produced in three
processes: quasiparticles incident from the superconductor are
transmitted into the normal metal with the probability $(1-A-B)$;
electrons are reflected from the interface with probability $B$; and
holes are Andreev-reflected as electrons with probability $A$. (The
latter process can be described in more direct terms as tunneling
of Cooper pairs out of the superconductor.) Thus, the energy
distribution $f^-(\epsilon)$ of electrons moving into the normal
metal is:
\begin{equation}
f^-(\epsilon) = A(\epsilon)(1-f^+(-\epsilon)) +
B(\epsilon)f^+(\epsilon)+(1-A(\epsilon)-B(\epsilon)) f(\epsilon)
\, . \label{2*} \end{equation}

 From the thermodynamic relation $dU=dQ+\mu dN$ applied to electron
gas in the normal metal the heat current $j$ flowing from the
normal metal into the superconductor in one (spin-degenerate)
transverse mode can be calculated as
\begin{equation}
j= \frac{1}{\pi \hbar} \int d \epsilon (\epsilon -eV)
(f^+(\epsilon)-f^-(\epsilon)) \, .
\label{3} \end{equation}
It is straightforward to check that when deviations of electron
energy distribution from equilibrium in each electrode are
small, eq.\ (\ref{3}) is equivalent to another frequently
used expression for the heat current, $j=Th$, where $T$ is the
temperature of the electrode and $h$ is the entropy flow.

The last step in the calculation is a summation over transverse
modes. We do this summation assuming that the electron motion
in the junction is quasiclassical, and that NS interface is smooth
and conserves momentum along it. Then the transmission probability
$D$ of the interface depends only on the energy $\epsilon_{\bot}$ of
electron motion across the interface, and can be taken to be $D(
\epsilon_{\bot})=D_0\exp \{ (\epsilon_{\bot}-\mu)/\epsilon_{tr} \}$.
In this expression, $D_0$ is the transparency at Fermi energy, and
$\epsilon_{tr}$ is an energy scale associated with electron
motion under the barrier, which is assumed to be much larger than
the superconducting energy gap. Under these assumptions summation
over transverse modes (i.e., over angles of incidence on the
interface) gives for the total heat current $J$:
\begin{equation}
J=\frac{N}{\pi \hbar} \int_0^{D_0} \frac{dD}{D} \int d
\epsilon (\epsilon -eV) (f^+(\epsilon)-f^-(\epsilon,D)) \, ,
\label{4} \end{equation}
where $N=Sm\epsilon_{tr} /\pi \hbar^2$ is the effective
number of transverse modes, $m$ is electron mass, and $S$ is
the junction area.

Taking the limit $D\rightarrow 0$ in eqs.\ (\ref{1}) and (\ref{2*}),
one can see that for $D_0 \rightarrow 0$ eq.\ (\ref{4}) for the
heat current reduces to the form given by the tunnel Hamiltonian
approach:
\begin{equation}
J=\frac{1}{e^{2} R_{T}} \int_{-\infty}^{+\infty} d\epsilon
\Theta (\epsilon^2 -\Delta^2) \frac{\mid \! \epsilon \! \mid
(\epsilon -eV)}{\sqrt{\epsilon^{2} -\Delta^{2}}}
[f(\epsilon-eV)-f(\epsilon) ]  \, ,
\label{5} \end{equation}
where $R_T$ is the normal-state tunneling resistance of the barrier,
$R_T^{-1} = Ne^2D_0 /\pi \hbar$.

Figure 1 shows the heat current as a function of the bias
voltage $V$ across the contact for different transparencies $D_0$
calculated numerically from eqs.\ (\ref{1}), (\ref{2*}), and
(\ref{4}). The curves illustrate how increasing transparency of the
barrier suppresses heat transfer out of the normal metal. We see
that for each transparency there is an optimal bias voltage which
maximizes the heat current $J$. The inset in Fig.\ 1 shows
$J$ at the optimal bias voltage in the tunneling limit calculated
from eq.\ (\ref{5}). The heat current $J$ in this limit is maximum
at $T\simeq 0.3 \Delta$ and decreases both at small and large
temperatures. One can check that at $T\ll \Delta$ the heat current
decreases as $(T/\Delta)^{3/2}$.

In Figure 2 we plot the heat current per one transverse mode at the
optimum bias voltage as a function of
barrier transparency $D_0$. At small transparencies the heat current
increases linearly with transparency, indicating that we are in the
tunneling limit where electron transport is dominated by the
single-particle tunneling. However, at larger transparencies the
heat current starts to decrease with transparency due to
increasing contribution to transport from the two-particle
tunneling. At the transition point between the two regimes the heat
flow out of the normal metal is maximized. A crude estimate of the
transparency which corresponds to this transition at low temperatures
can be obtained if we
compare the amount of heat per transverse mode generated by the
two-particle tunneling in the normal electrode,
$\simeq (D_0\Delta)^2/\hbar$, with the heat flow out of this
electrode in the tunneling regime. As was noted above, the latter
can be estimated as $(T^3\Delta)^{1/2}D_0/\hbar$ (see also inset in
Fig.\ 1). From these estimates we see that at $T\ll \Delta$ the
optimal transparency scales as $(T/\Delta)^{3/2}$. This conclusion
is in qualitative agreement with Fig.\ 2.

\section{Disordered contacts} \label{sec:dis}

\vspace{1ex}

In this section we consider the same model of a short NIS
constriction, but assume that the constriction dimensions $d$
are much larger than the elastic mean free path $\ell$ in the
electrodes. A convenient way to describe such a disordered
constriction is provided by the quasiclassical kinetic equations
for non-equilibrium Green's function $\check{G}$ of the electrodes
\cite{Eilen,Us,LaOv}. Green's function $\check{G}$ is a
triangular matrix in the Keldysh space:
\begin{equation}
\check{G} = \left( \begin{array}{cc} \hat{G}^{R} & \hat{G}^{K} \\
0 & \hat{G}^{A} \end{array} \right) \, ,
\label{21} \end{equation}
where the advanced $\hat{G}^{A}$, retarded $\hat{G}^{R}$, and Keldysh
component $\hat{G}^{K}$ are $2\times 2$ matrices in the electron-hole
(Nambu) space. The retarded and advanced functions carry information
about the excitation spectrum of the system, while the Keldysh
function describes the distribution of quasiparticles.

For small constrictions, $d \ll \xi$, we can neglect all but the
gradient term in the equation for Green's function in the vicinity
of the constriction, so that the equation is reduced to the
diffusion equation \cite{Eilen,Us,LaOv}:
\begin{equation}
\vec{\nabla} ({\cal D} \check{G} \vec{\nabla} \check{G})=0 \, ,
\label{22} \end{equation}
with the diffusion coefficient ${\cal D}=\frac{1}{3} v_{F} \ell$,
where $v_F$ is the Fermi velocity. Diffusion equation in the
electrodes should be complemented by the boundary condition at the
NS interface \cite{Za,KuL}:
\begin{equation}
c \vec{n} \check{G}  \vec{\nabla} \check{G}  =
[ \check{G}, \check{G}_{S}] \, .
\label{23} \end{equation}
In this expression, $\vec{n}$ is the vector normal to the NS
interface, and $\check{G}$, $\check{G}_S$ are the Green's
functions on the normal and superconducting sides of the
interface, respectively. The coefficient $c$ describes the interface
transparency and can be written as $2\sigma/g$, where
$\sigma=2e^2 \nu {\cal D}$ is the conductivity
of the normal metal, and $g= e^2 \nu \langle \vec{n} \vec{v}_F
D/R \rangle$ is the normal-state conductance per unit area of the
interface. Here $D$ and $R$ are the transmission and
reflection probabilities
of the interface, $\nu$ is the density of states at the Fermi
level, and $\langle ... \rangle$ denotes averaging over angles
of incidence on the interface.

On a quantitative level, properties  of the constriction
depend on its specific geometry. Below we consider a simple
one-dimensional (1D) model shown schematically in Fig.\ 3.
In this model the constriction is represented as a 1D normal
conductor of length $d$, cross-section area $A$ and resistance
$R_N=d/(A\sigma)$, and it is assumed that the contribution of the
bulk regions to the total resistance of the structure is much
smaller than $R_N$. In this case we can neglect variations of
the order parameter $\Delta$ and Green's function $\check{G}_{S}$
in the superconducting electrode, and assume that $\Delta$ and
$\check{G}_{S}$ are equal to their equilibrium values up to the
NS boundary:
\begin{equation}
G^{R(A)}_S = \left( \begin{array}{cc} g^{R(A)} & f^{R(A)} \\
-f^{R(A)} & -g^{R(A)} \end {array} \right)
= \frac{1}{\sqrt{ (\epsilon \pm i0)^{2} - \Delta^{2}}}
\left( \begin{array}{cc} \epsilon & \Delta \\
-\Delta & -\epsilon \end{array} \right) \, ,
\label{27} \end{equation}
\begin{equation}
G^K_S =G^R_S (\epsilon) n(\epsilon)-n(\epsilon)G^A_S(\epsilon) \, ,
\;\;\;\; n(\epsilon) = \tanh (\frac{\epsilon}{2T} ) \, .
\label{28} \end{equation}
In the 1D constriction (Fig.\ 3) the Green's functions depend only
on the coordinate $x$ along the constriction, so that the diffusion
equation becomes:
\begin{equation}
\frac{\partial}{\partial x} (\check{G}
\frac{\partial}{\partial x} \check{G} ) =0 \, .
\label{24} \end{equation}

Normalization condition on $\check{G}$, $\check{G}^{2}=
\check{1}$, \cite{LaOv,shel} implies that the retarded and advanced
components can be parameterized as follows \cite{vol}:
\begin{equation}
\hat{G}^{R(A)}(\epsilon)=\pm ( \hat{\sigma}_{z} \cosh[U^{R(A)}
(\epsilon)] + i \hat{\sigma}_{y} \sinh[U^{R(A)}(\epsilon)]) \, ,
\label{25} \end{equation}
where $\sigma's$ are Pauli matrices, and upper and lower signs
are for $\hat{G}^R$ and $\hat{G}^A$ respectively. The fact that
the retarded and advanced functions are nondiagonal indicates
the proximity effect in the normal region. $U^{R(A)}(\epsilon)$
determines the local density of states as a function of the
energy $\epsilon$. The parameterization (\ref{25}) allows us to
think about it as an imaginary polar angle in the 2-dimensional
space of $(\hat{\sigma}_{z}, \hat{\sigma}_{y} )$. The magnitude
of the angle corresponds to the degree of ``proximization'' of
the normal region.

The Green's functions should satisfy the equilibrium boundary
condition at the normal end of the constriction ($x=-d$):
$\hat{G}^{R(A)} = \pm \hat{\sigma}_{z}$, i.e., $U^{R(A)} = 0$.
Equation (\ref{24}) with this boundary condition determines
$U^{R(A)}(x)$:
\begin{equation}
U^{R(A)} = a^{R(A)}(1+\frac{x}{d}) \, ,
\label{25*} \end{equation}
where $a^{R(A)}(\epsilon) \equiv U^{R(A)}(x=0,\epsilon)$.
Substituting eqs.\ (\ref{25}) and (\ref{25*}) into the boundary
condition (\ref{23}) we reduce it to a transcendental equation:
\begin{equation}
\pm \frac{R_{T}}{R_{N}} a^{R(A)} = f^{R(A)} \cosh a^{R(A)}-
g^{R(A)} \sinh a^{R(A)} \, .
\label{26} \end{equation}
Here $g^{R(A)}$ and $f^{R(A)}$ are components of the equilibrium
Green's functions (\ref{27}) of the superconductor, and $R_T$ is
the normal-state tunneling resistance of the NS interface,
$R_T^{-1} =gA$.

The next step is to find the Keldysh function $\hat{G}^K$
which can be represented as \cite{LaOv}:
\begin{equation}
\hat{G}^K =\hat{G}^R (\epsilon)\hat{h}(\epsilon)-\hat{h}(\epsilon)
\hat{G}^A(\epsilon) \, , \;\;\;\; \hat{h}(\epsilon) =
f_{1}(\epsilon) \hat{1} +f_{z}(\epsilon) \hat{\sigma}_{z} \, .
\label{30} \end{equation}
Equation (\ref{24}) with the retarded and advanced components
(\ref{25}) gives the following equations for the distribution
functions $f_{1(z)}$ (\ref{30}):
\begin{equation}
(1+\cosh (U^{R} \mp U^{A})) \frac{\partial f_{1(z)} }{\partial x}
=\mbox{const} \equiv B_{1(z)}/d \, .
\label{31} \end{equation}
Combining these equations with the boundary condition (\ref{23})
at $x=0$ and equilibrium boundary condition at $x=-d$:
$f_{1,(z)}(-d,\epsilon) = ( n(\epsilon-eV) \pm n(\epsilon+eV))/2$,
we find the functions $B_{1,z}$ which determine currents across
the NS interface:
\begin{equation}
B_1(\epsilon) = \frac{A_1(\epsilon)}{D_1}(2n(\epsilon)
-n(\epsilon-eV)-n(\epsilon+eV)) \, , \;\;\; B_z(\epsilon)
= \frac{A_z(\epsilon) }{D_z} (n(\epsilon-eV)-n(\epsilon+eV)) \, ,
\label{32} \end{equation}
where
\[ A_{1(z)} = (g^R -g^A) (\cosh a^R +\cosh a^A)-(f^R \mp f^A)
(\sinh a^R \pm \sinh a^A) \, , \]
\[ D_{1(z)} = 4 [ \frac{R_T}{R_N} +A_{1(z)}(\epsilon)
\frac{\tanh[(a^R \mp a^A)/2]}{2 (a^R \mp a^A) } ] \, . \]

Equations (\ref{26}) and (\ref{32}) enable us to find electric
current and heat current in the NS junction. The electric current
can be written as follows:
\begin{equation}
I = \frac{e \nu {\cal D}A}{2} \int_{-\infty}^{+\infty}
d\epsilon \; \mbox{tr} \{ \hat{\sigma}_{z} [\check{G} \vec{\nabla}
\check{G}]^K (\epsilon) \} = \frac{1}{2eR_N} \int_{-\infty}
^{+\infty} d\epsilon B_z (\epsilon) \, .
\label{33} \end{equation}

Figure 4 shows the current $I$ (\ref{33}) as a function of the
bias voltage $V$ at several ratios of the resistances $R_T/R_N$
and vanishing temperature. At large tunnel resistances $R_T$
the $I-V$ characteristic exhibits the gap at $V<\Delta/e$
due to vanishing density of states in the superconductor, and
associated singularity at $V=\Delta/e$. At smaller $R_T$ the
gap is closed by the increasing contribution to the current
from Andreev reflection, but the $I-V$ characteristics still
has the singularity (logarithmic divergence of the
differential conductance) at $V=\Delta/e$. Transition between the
two regimes is clearly visible in the zero-bias linear conductance
shown in the inset in Fig.\ 4. At
$R_T<R_N$ the conductance with a good accuracy equals simply
$(R_T+R_N)^{-1}$, while at $R_T>R_N$ it decreases as $R_N/R_T^2$.

Recalling the definition of the heat current used in the previous
section and eq.\ (\ref{33}) for the electric current, we can write
the following expression for the heat flow out of the normal
electrode in disordered NIS junction:
\begin{equation}
J = \frac{1}{2e^{2} R_{N}} \int_{-\infty}^{+\infty} d\epsilon
(\epsilon-eV) (B_{1}(\epsilon) + B_{z}(\epsilon))\, .
\label{34} \end{equation}
As can be seen from eqs.\ (\ref{26}) and (\ref{32}), the currents
$B_{1,z}$ have the property $B_{1(z)}(-\epsilon) = \mp
B_{1(z)}(\epsilon)$. With these relations, eq.\ (\ref{34}) can be
rewritten as follows:
\begin{equation}
J = -IV+ \frac{1}{e^{2} R_{N}} \int_0^{\infty} d\epsilon
\epsilon B_{1}(\epsilon) \, .
\label{35} \end{equation}
At $R_T \gg R_N$, both eq.\ (\ref{33}) for the electric current
and eqs.\ (\ref{34}), (\ref{35}) for the heat current are
reduced to the form given by the tunnel Hamiltonian approach. In
particular, the heat current $J$ is again given by eq.\ (\ref{5}).

Some results of the numerical calculation of the heat current
from eqs.\ (\ref{26}), (\ref{32}), and (\ref{35}) are shown in
Figs.\ 5 and 6. We see that the properties of the heat current
in disordered contacts are qualitatively similar to those for
clean contacts. The heat current density grows with increasing
transparency of the tunneling barrier in the regime of small
transparencies (large tunnel resistances $R_T$) and is gradually
suppressed at large
transparencies. The main difference with the clean case is that
the scale of transparencies (in particular, the optimum
transparency) is shifted downwards by a small factor $\ell/d$,
which decreases with increasing resistance of the normal
electrode. The physical reason for this shift is that disorder
in the normal electrode enhances contribution of Andreev
reflection to transport and therefore suppresses the heat flow
out of the normal metal.

\section{Conclusion}
\label{sec:concl}

In conclusion, we have calculated the heat current in clean and
disordered NIS microcontacts which is caused by electric current
flow across the NS interface. The mechanism of the heat transfer
is analogous to that of the Peltier effect in normal
metal/semiconductor structures.
Results for clean NIS junctions are obtained by solving the
Bogolyubov-de Gennes equations. Disordered junctions are described
with the quasiclassical equations for non-equilibrium Green's
functions of the electrodes. In both
cases the heat current density exhibits non-monotonic dependence on
the interface transparency, increasing at small transparencies and
decreasing at large transparencies. The transition between the two
regimes takes place at the transparency which is determined by the
interplay of single-particle tunneling and Andreev reflection.

At intermediate temperatures, $T\simeq \Delta$, this transition
occurs at the transparencies which are larger than the barrier
transparencies of the typical tunnel junctions between good metals.
For instance, even quite small specific tunneling resistance on the
order of 10 Ohm$\times \mu$m$^2$ corresponds to barrier transparency
$\simeq 10^{-4}$. However, as it was demonstrated above, as
temperature decreases, the transition point moves rapidly towards
smaller transparencies, and at low temperatures finite barrier
transparency can pose an important limitation on the refrigeration
power of NIS junctions. As follows from our estimates in Sec.\ 2,
in the clean case transparency-related limitation should become
important at $T\simeq 0.01 \Delta$ in junctions with specific
resistance 10 Ohm$\times \mu$m$^2$ or less.

\vspace{1ex}

The authors gratefully acknowledge discussions with K. Likharev.
This work was supported in part by ONR grant \# N00014-95-1-0762.

\figure{ Figure 1.
Heat current $J$ in the clean NIS contact versus bias
voltage $V$ for several transparencies $D_0$ of
the insulator barrier calculated from eq.\ (\protect \ref{4}).
 From top to bottom, $D_0=$ 0 (tunneling limit),
0.03, 0.1, and 0.2. The inset shows the heat current
in the tunneling limit calculated for the optimum bias voltage as a
function of temperature; at low temperatures $J\propto
(T/\Delta)^{3/2}$. }

\figure{ Figure 2.
The maximum heat current density in the clean NIS contact
as a function of transparency $D_0$ of the insulator barrier for
several temperatures. For discussion see text. }

\figure{ Figure 3.  Schematic diagram of
disordered NIS contact. Darker region shows the quasi-1D
constriction of length $d$ which determines the resistance
$R_N$ of the normal electrode. An insulator barrier with
normal-state resistance $R_T$ is placed at the NS interface
in the constriction. }

\figure{ Figure 4.  $I-V$ characteristics of
disordered NIS contact at zero temperature and several
values of the resistance ratio $R_T/R_N$. From top to bottom,
$R_T/R_N=\, 0,\, 1,\, 3,\, 10,\, 100$. The inset shows the
linear conductance $G=dI/dV\mid_{V=0}$ as a function of
$R_T/R_N$. The curves illustrate the transition between the
tunneling regime ($R_T>R_N$) characterized by the gap at
$V<\Delta/e$ and ``metallic'' regime ($R_T<R_N$)
characterized by large subgap conductance and excess current
at $V\gg \Delta/e$. }

\figure{ Figure 5.
Heat current $J$ in the disordered NIS contact versus bias
voltage $V$ for several ratios of the normal electrode
resistance $R_N$ to the resistance $R_T$ of the insulator barrier.
As in the clean contacts, the heat current is maximized at
$V\simeq \Delta/e$.  }

\figure{ Figure 6.
The heat current density in the disordered NIS contact
calculated at the optimum bias voltage as a function of the
insulator barrier resistance $R_T$. The curves are similar to those
in Fig.\ 2 for clean contacts. Note, however, that the scale of the
$x$-axis, $R_N/R_T$, for disordered contacts corresponds to much
smaller transparencies of the tunnel barrier than for clean
contacts ($R_N/R_T \simeq (d/\ell) D_0 \gg D_0$).  }


\begin{references}
\bibitem{b1} M. Nahum, T.M. Eiles, and J.M. Martinis,
Appl.\ Phys.\ Lett. {\bf 65}, 3123 (1994).
\bibitem{b2} M.G. Blamire, E.C.G. Kirk, J.E. Evetts, and T.M.
Klapwijk, Phys.\ Rev.\ Lett. {\bf 66}, 220 (1991); D.R. Helsinga
and T.M. Klapwijk, Phys.\ Rev.\ B {\bf 47}, 5157 (1993).
\bibitem{b3} A.V. Zaitsev, JETP Lett. {\bf 55}, 66 (1992).
\bibitem{b4} K. Seeger, {\em Semiconductor Physics}
(Springer, NY, 1991), Ch.\ 4.
\bibitem{b5} A.F. Volkov, A.V. Zaitzev, and T.M Klapwijk,
Physica C {\bf 210}, 21 (1993).
\bibitem{b6} C.W.J. Beenakker, B. Rejaei, and J.A. Melsen,
Phys.\ Rev.\ Lett. {\bf 72}, 2470 (1994).
\bibitem{b7} Yu.V. Nazarov, Phys.\ Rev.\ Lett. {\bf 73},
1420 (1994).
\bibitem{b8} G.E. Blonder, M. Tinkham, and T.M. Klapwijk,
Phys.\ Rev.\ B {\bf 25}, 4515 (1982).
\bibitem{b9} C.W.J. Beenakker, in: {\em Mesoscopic Quantum
Physics}, E. Akkermans, G. Montambaux, and J.-L. Pichard, eds.
(North-Holland, to be published).
\bibitem{Eilen} G. Eilenberger, Z.\ Phys. {\bf 214}, 185 (1968).
\bibitem{Us} K. Usadell, Phys.\ Rev.\ Lett. {\bf 25}, 507 (1970).
\bibitem{LaOv} A.I. Larkin and Yu.N. Ovchinnikov, Sov.\ Phys.\
JETP {\bf 41}, 960 (1975); {\bf 46}, 155 (1977).
\bibitem{Za} A.V. Zaitsev, Sov.\ Phys.\ JETP {\bf 59}, 1015 (1984).
\bibitem{KuL} M.Yu. Kuprianov and V.F. Lukichev, Sov.\ Phys.\ JETP
{\bf 67}, 1163 (1988).
\bibitem{shel} A.L. Shelankov, J. Low Temp.\ Phys. {\bf 60}, 29
(1985).
\bibitem{vol} A.F. Volkov, JETP Lett. {\bf 55}, 474 (1992).
\end{references}
\end{document}